\documentclass[twocolumn,amsmath,amssymb,aps,pra,superscriptaddress,showpacs,tightenlines]{revtex4}
\usepackage[latin9]{inputenc}
\usepackage{epsfig}
\usepackage{bm}
\usepackage{amssymb}
\usepackage{graphicx}
\usepackage{epstopdf}
\usepackage{amsmath}
\usepackage{natbib}
\usepackage{hyperref}

\usepackage{color}
\providecommand{\U}[1]{\protect\rule{.1in}{.1in}}
\begin{document}
	\title{Hybrid Interference Induced Flat Band Localization in Bipartite Optomechanical Lattices}
	\author{Liang-Liang Wan}
	\affiliation{School of physics, Huazhong university of science and technology, Wuhan 430074, China}
	\author{Xin-You L\"u}
	\email{xinyoulu@hust.edu.cn}
	\affiliation{School of physics, Huazhong university of science and technology, Wuhan 430074, China}
	\author{Jin-Hua Gao}
	\email{jinhua@hust.edu.cn}
	\affiliation{School of physics, Huazhong university of science and technology, Wuhan 430074, China}
	\author{Ying Wu}
	\email{yingwu2@126.com}
	\affiliation{School of physics, Huazhong university of science and technology, Wuhan 430074, China}
	\date{\today}

\begin{abstract}
The flat band localization, as an important phenomenon in solid state physics, is fundamentally interesting in the exploration of exotic
ground property of many-body system. Here we demonstrate the appearance of a flat band in a general bipartite optomechanical lattice, which could have one or two dimensional framework.
Physically, it is induced by the hybrid interference between the photon and phonon modes in optomechanical lattice, which is quite different from the destructive interference resulted from the
special geometry structure in the normal lattice (e.g., Lieb lattice). Moreover, this novel flat band is controllable and features a special local density of states (LDOS) pattern, which makes it is detectable in experiments.
This work offers an alternative approach to control the flat band localization with optomechanical interaction, which may substantially advance the fields of cavity optomechanics and solid state physics.
\end{abstract}

\maketitle

\section{Introduction}

The electron localization in a crystal is an important phenomenon in solid state physics, which relates to many fundamental problems, e.g., the metal-insulator transition.  Normally, the localization phenomenon is due to the presence of disorder, i.e., the celebrated Anderson localization~\cite{Anderson1958}. But in some special crystal lattices, electrons can be localized without any disorder and form a completely flat band in the whole Brillouin zone, the reason of which is the destructive wave interference resulted from the lattice geometry~\cite{Tasaki2008,BERGHOLTZ2013,LiuZheng2014}. This is just the flat band localization. Lieb, Kagome, Diamond, stub, and sawtooth lattices are some examples of the flat band lattices, and some general methods are proposed to design more lattice structures with flat bands~\cite{LiuZheng2013}. That the flat band electrons are of special interest is because that, due to the quenched kinetic energy, tiny interaction can induce some exotic correlated ground states, such as  ferromagnetism~\cite{Tasaki2008,Lieb1989,Mielke1991,Shen1994}, superconductivity~\cite{Wu2007}, and  Wigner crystals~\cite{Miyahara2007,Julku2016,Kopnin2011}. However, though the flat band electrons have been intensively studied in last three decades, it has not been experimentally confirmed in natural materials due to the complexity of real materials.

Most recently, a essential progress about the flat band lattice has been achieved in artificial quantum lattice systems, where flat band lattices have been realized in experiment, and flat band localization is observed as well~\cite{Jacqmin2014,Mukherjee2015,Vicencio2015}. For example, it is reported that the Lieb lattice has been realized in various quantum systems, such as photonic crystals~\cite{Aspuru-Guzik2012,Carusotto2013,Mukherjee2015,Vicencio2015}, cold atoms~\cite{Taie2015}, artificial electron lattice on metal surface~\cite{Gomes2012,Wang2014prl}. Interestingly, the flat band localization are clearly demonstrated in both the  photonic crystals~\cite{Mukherjee2015,Vicencio2015,Yang2016} and the artificial electron lattice on metal surface~\cite{Drost2017,Slot2017,Qiu2016}, while the flat band superconductivity is realized in the cold atom system~\cite{Taie2015}.

The cavity optomechanical system is a hybrid artificial quantum system combining the optical and mechanical modes, which has developed rapidly in the last decade~\cite{Kippenberg2008,Marquardt2009,Aspelmeyer2012,Meystre2013,Aspelmeyer2014,Sun2015,Xiong2015a,Lue2015a, Lue2015,Lue2015prlb,Clark2017}.
In particular, the realization of optomechanical crystals~\cite{Eichenfield2009,Safavi-Naeini2010,Chan2011,Gavartin2011,Safavi-Naeini2014} offers an alternative platform of investigating both light and sound propagation in optomechanical arrays. For example, slow light~\cite{Chang2011}, photon propagation~\cite{Chen2014}, optical solitons~\cite{Gan2016} and polarizer~\cite{Xiong2016} have been proposed in the optomechanical arrays. Moreover, the optomechanical arrays also could exhibit the interesting quantum many-body physics~\cite{Tomadin2012, Ludwig2013}, e.g., the non-trivial topological phases of light and sound have been demonstrated~\cite{Peano2015}.
A nature question is whether the optomechanical interaction will influence the flat band localization significantly.

Here, we investigate the flat band localization in the bipartite optomechanical lattice. Our main finding is that, instead of the destructive interference resulted from the lattice geometry, the hybrid photon-phonon-interference in optomechanical lattice can also induce the flat band localization.  An immediate consequence is that, in the optomechanical lattice, the structure of the flat band lattice, as well as the corresponding band structures, can be different from that in other lattices. This actually reflects the quantum interference characteristic of the optomechanical lattice.

\begin{figure}
	\centering
	\includegraphics[width = 7.5cm]{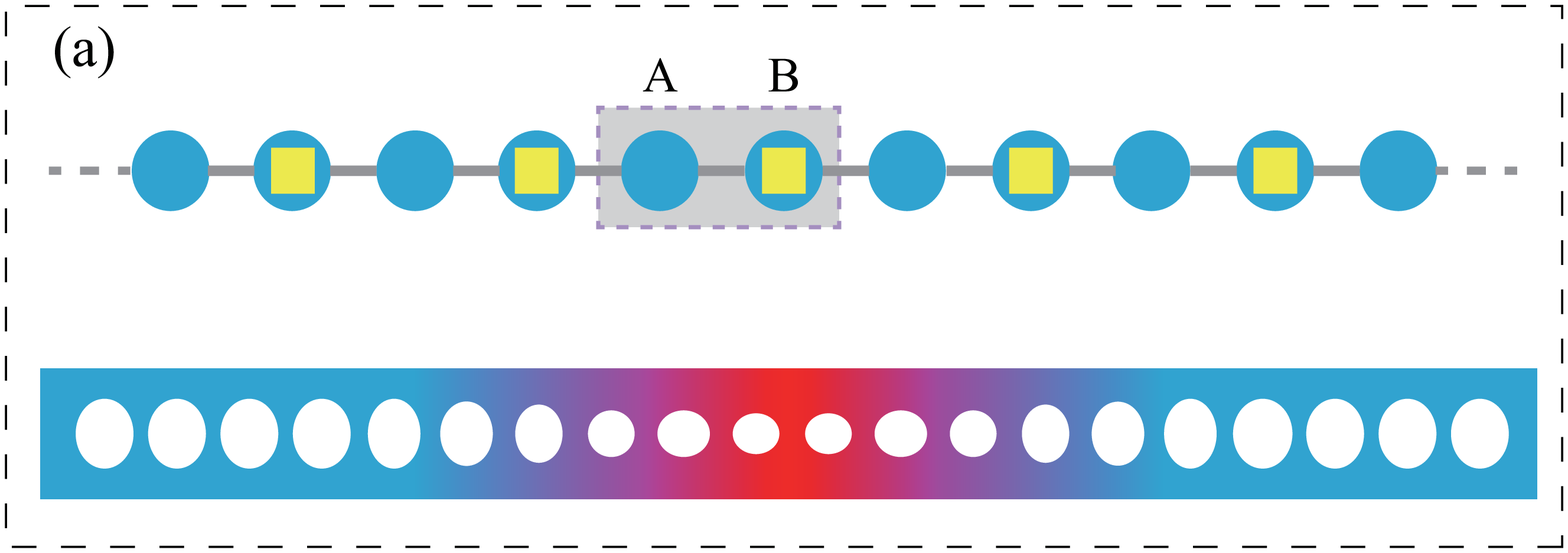}
	\includegraphics[width = 7.5cm]{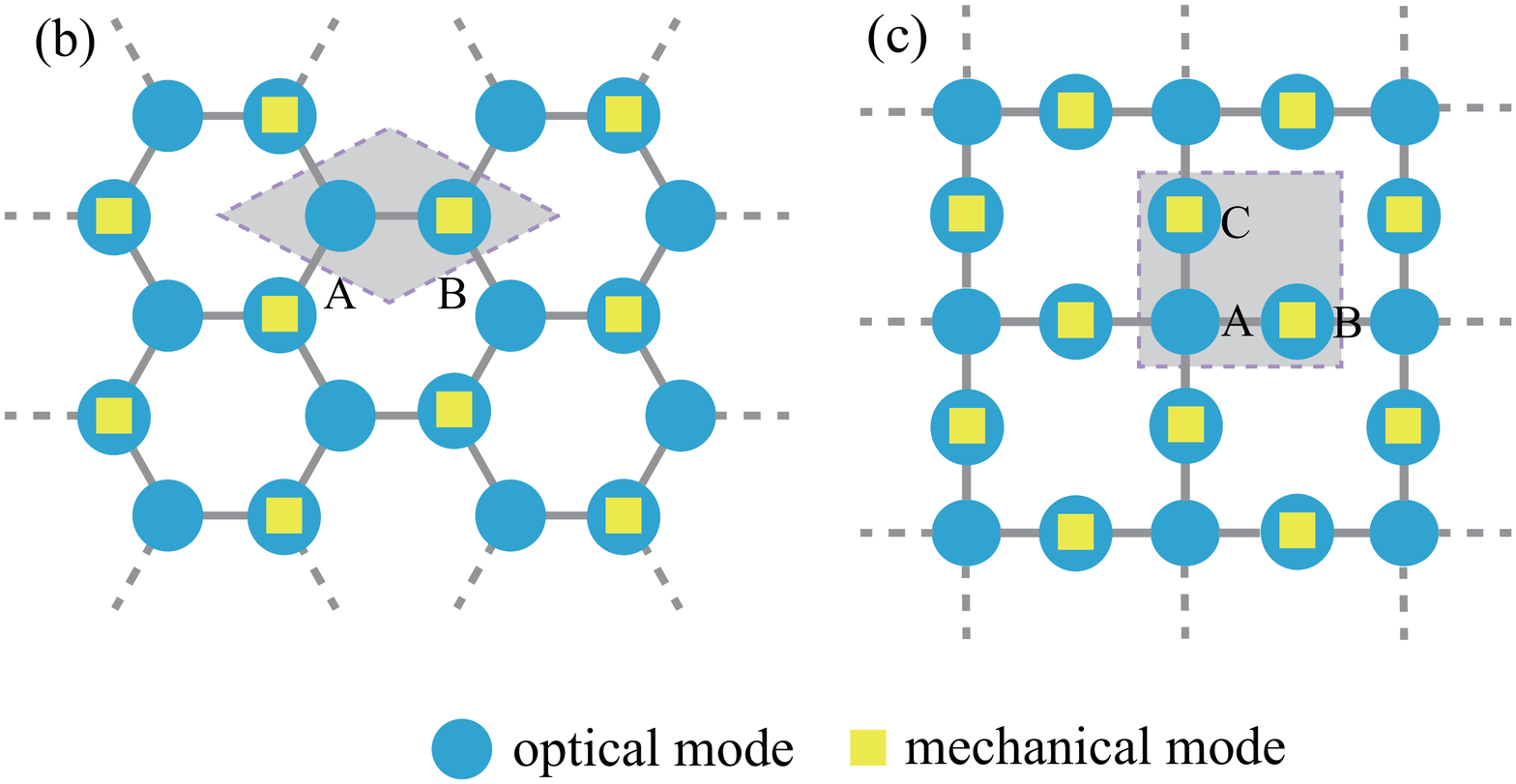}\\
	\caption{(Color online) The sketch of hybrid bipartite optomechanical lattices with (a) one dimension framework, and two dimension (b) honeycomb, (c) Lieb frameworks. They both are bipartite lattices, which can be separated into two sublattices, denoted $A$ and $B,C$. The sublattice A only has the optical mode (pure optical site), while sublattice $B$ or $C$ has coupled optical and mechanical modes (optomechanical site). The lower illustration in (a) denotes the implementation of one dimensional hybrid lattice with the optomechanical crystals.}
	\label{fig1}
\end{figure}

Concretely, we propose the model of hybrid bipartite optomechanical lattice with one or two dimensional framework, whose unit cell consists of one optical sublattice and one optomechanical sublattice.
A novel flat band is found, which is resulted from the hybrid interference between phonon and photon modes, and does not exist in other artificial quantum lattice systems.
More interestingly, this new flat band corresponds to a special photon-phonon-localization pattern, i.e., photons are only localized on the optical sublattice, while phonons are only localized on the optomechanical sublattice. In the previous lattices which exist flat-band localization, excitations~(photons or phonons) are only localized on one sublattice in comparison.
This property can be used to identify this new flat band state in experiment. Furthermore, this new flat band together its special localization pattern is controllable by the driving laser applied into the bipartite optomechanical lattices, which is an obvious advantage compared with the previous flat band localization induced by the lattice geometry. Note that, for the case of two dimensional optomechanical lattice, we choose the optomechanical honeycomb and Lieb lattice as the examples. In the optomechanical Lieb lattice, we find three flat bands. We demonstrate that the three flat bands have two different origins. The middle one is the new flat band and it is because of the hybrid interference, and the others are due to the lattice geometry. The two kinds of flat band also have distinct photon-phonon-localization patterns.

\section{model and methods}

We consider a bipartite optomechanical lattice, whose unit cell consists of an optical sublattice (i.e., site A) and an optomechanical sublattice (i.e., site B).
As shown in Fig.\, 1, this bipartite optomechanical lattice can have one or two dimensional framework.
Applying a strong laser with frequency $\omega_L$ on the bipartite optomechanical lattice, the system Hamiltonian in a frame rotating with $\omega_L$ reads~\cite{Pace1993,Kippenberg2007,Vitali2007,Marquardt2007}
\begin{eqnarray}
H = \sum_{i\alpha,\beta}(\Delta a_{i\alpha}^\dagger a_{i\alpha} + \Omega b_{i\beta}^\dagger b_{i\beta})+H_{\rm{int}},
\\
H_{\rm{int}} = \sum_{i\beta}g(a_{i\beta}^\dagger b_{i\beta} + a_{i\beta} b_{i\beta}^\dagger)+\sum_{\langle i\alpha,j\alpha'\rangle}Ja_{i\alpha}^\dagger a_{j\alpha'},
\end{eqnarray}
where $a$~($a^\dagger$) and $b$~($b^\dagger$) are the annihilation~(creation) operators of the optical and mechanical modes, respectively.
The joint index $(i,\alpha)$ and $(i,\beta)$ contain the subindexes $\alpha$ and $\beta$. The Hamiltonian can represent a one or two dimensional lattice by properly defining $i$, $\alpha$ and $\beta$. For the one dimension (1D) lattice shown in Fig.\,1(a), $\alpha = A, B$, and $\beta = B$, denoting A-sites and B-sites, respectively.
And $i$ indexes the unit cell. In words, for the case each site has a localized optical mode, which is evanescently coupled to the optical modes at adjacent sites. Its unit cell can be separated into two sublattice, the optical sublattice, site A and the optomechanical sublattice, site B. In the optomechanical sublattice, a localized mechanical mode couples to the optical mode in the same site.
Extending to the case of two dimension (2D), two kinds of bipartite lattices are considered here, i.e., the hybrid honeycomb lattice and Lieb lattice, shown in Fig.\,1(b,c).
Now $i$ denotes the unit cell, $\alpha = A,B$, $\beta = B$ correspond to the optomechanical honeycomb lattice.
And $\alpha = A,B,C$ and $\beta = B,C$ correspond to the optomechanical Lieb lattice, denoting A-sites, B-sites and C-sites. Similar as the 1D case, the unit cell includes an optical sublattice and an optomechanical sublattice. Then the nearest neighbor optical hopping with strength $J$ is considered and it is denoted by $\langle i,\alpha,j,\alpha'\rangle$.
The frequency detuning $\Delta \approx \omega_c-\omega_L$ with the cavity frequency $\omega_c$, and $g$ is the linearized optomechanical interaction strength, which is much smaller than the mechanical frequency $\Omega$.

In principal, the proposed hybrid bipartite optomechanical lattice is general and could be implemented in cavity (or circuit) QED system in the optical (or microwave) frequency range~\cite{Houck2012}.
As shown in Fig.\,1(a), the 1D optomechanical lattice could be realized in the optomechanical crystals~\cite{Eichenfield2009,Safavi-Naeini2010,Chan2011,Gavartin2011,Safavi-Naeini2014}. The defects are generated by a appropriate local modification of the pattern of holes, which localizes the optical and mechanical modes on the crystals. The accessible system parameters for our model could be $\lambda = 1,550$ nm, $\Omega/2\pi = 3.75$ GHz, $\kappa/2\pi \approx 900$ MHz, $\gamma/2\pi \approx 250$ KHz, and $g \approx \kappa$. Here $\lambda$, $g$ are the optical wavelength and linearized optomechanical coupling strength under the condition of strongly optical driving, respectively, and $\kappa$, $\gamma$ are the optical and mechanical decay rate.

Here we will examine the local density of states\,(LDOS) of lattice sites for both photon $\rho_{O}(\omega; j\alpha)$  and phonon $\rho_{M}(\omega; j\beta)$. In experiments, the LDOS of the photon at each site can be directly measured via a auxiliary probe laser. The photon and phonon LDOSes are formally defined as
\begin{eqnarray}
\rho_{O}(\omega; j\alpha) &= -2{\rm{Im}}G^R_{OO}(\omega;j\alpha,j\alpha), \nonumber
\\
\rho_{M}(\omega; j\beta) &= -2{\rm{Im}}G^R_{MM}(\omega;j\beta,j\beta),
\end{eqnarray}
where $G^R_{OO}(\omega;j\alpha,j\alpha)$ and $G^R_{MM}(\omega; j\beta,j\beta)$ are the retarded Green's function of photons and phonons in real space, respectively. And the definitions are
\begin{eqnarray}
G^R_{OO}(j\alpha,t;j'\alpha',t') &= -i\theta(t-t')\langle[a_{j\alpha}(t),a_{j'\alpha'}^\dagger(t')]\rangle, \nonumber
\\
G^R_{MM}(j\beta,t;j'\beta',t') &= -i\theta(t-t')\langle[b_{j\beta}(t),b_{j'\beta'}^\dagger(t')]\rangle.
\end{eqnarray}
To calculate the Green's functions, we start from the Heisenberg-Langevin equation of motion in momentum space,
\begin{equation}
i\partial_{t}{\psi} = M\psi +\xi,
\label{eq6}
\end{equation}
where $\psi$ is the vector of the photonic and phononic annihilation operators of lattices,  and $\xi$ is the vector of the noise operators of baths. Their specific formula depends on the lattice considered.
For example, $\psi = (a_{{\bf{k}}A}, a_{{\bf k}B}, b_{{\bf{k}}B})^T$ and $\xi = (a_{{\bf{k}}A}^{\rm{in}}, a_{{\bf{k}}B}^{\rm{in}}, b_{{\bf{k}}B}^{\rm{in}})^T$ for the 1D case we consider. In ${\bf k}$ space, the retarded Green's function $G^R({\bf k}; t, t') = -i\theta(t-t')\langle [\psi(t), \psi^\dagger(t')]\rangle$ satisfied
\begin{equation}
(i\partial_{t} -M)G^{R}(k;t,t') = \delta(t-t').
\end{equation}
Then the retarded Green's function can be obtained via Fourier transformation:
\begin{equation}
G^R(\omega;{\bf{k}}) = (\omega- M)^{-1}.
\label{Green-function}
\end{equation}
The diagonal components of $G^R(\omega;{\bf k})$ matrix give the photonic and phononic retarded Green's functions. With Fourier transformation, the photonic and phononic LDOSes can be expressed as:
\begin{eqnarray}
\rho_{O}(\omega;j\alpha) = -\frac{2}{N}\rm{Im}\sum_{\bf k} G^R_{OO}(\omega;{\bf k}\alpha),\nonumber
\\
\rho_{M}(\omega;j\beta) = -\frac{2}{N}\rm{Im}\sum_{\bf k} G^R_{MM}(\omega;{\bf k}\beta),
\label{eq8}
\end{eqnarray}
where $N$ is the number of unit cells of lattices.

\subsection{hybrid one-dimension optomechanical lattice}

\begin{figure}[h]
	\includegraphics[width = 8.5cm]{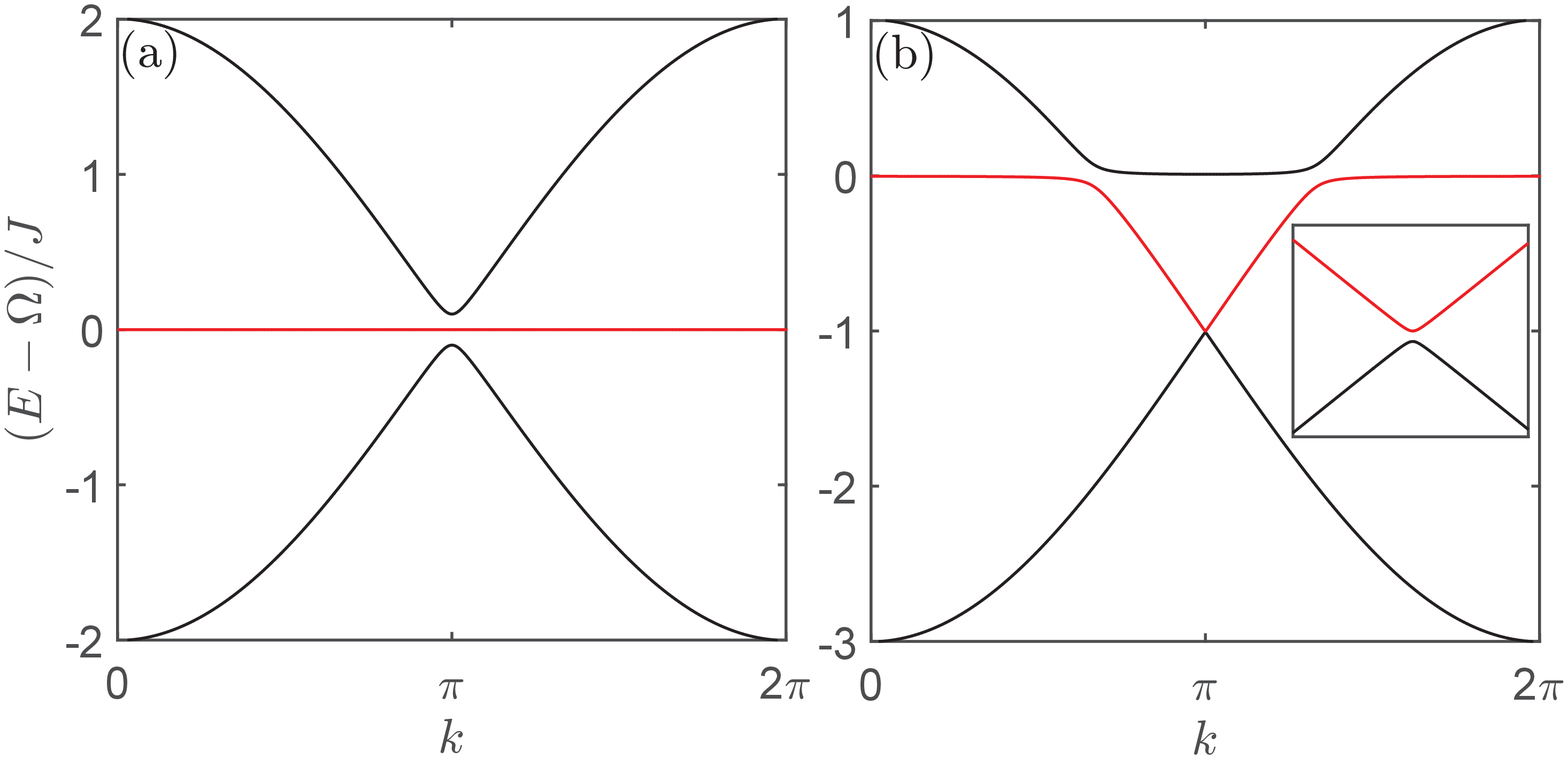}\\
	\caption{(Color online) (Color online) Energy structure of one dimensional optomechanical array lattice when (a) $\Delta = \Omega$ and (b) $\Delta =\Omega-J$. The system parameters are $g = 0.01J$.}
	\label{fig2}
\end{figure}

\begin{figure}[h!]
	\includegraphics[width = 8.5cm]{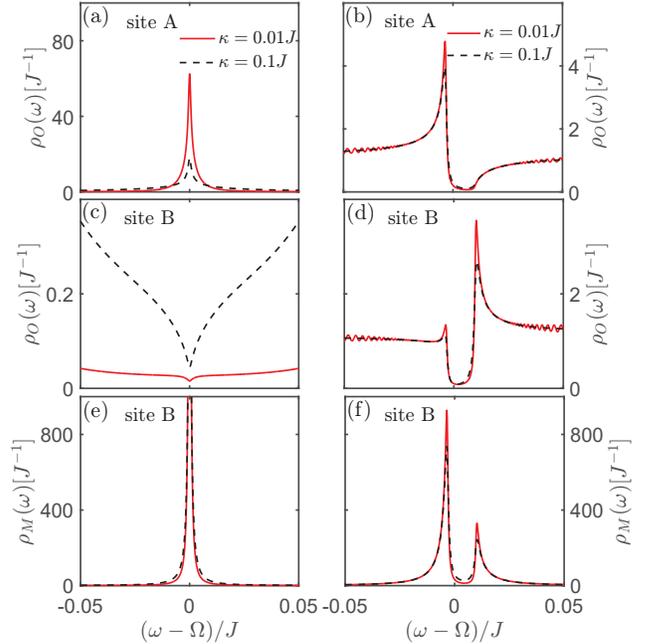}
	\caption{(Color online) (Color online) The LDOS of 1D array for different decay rate $\kappa$. The local photon or phonon DOS of sites A and B under the condition (a,c,e) $\Delta=\Omega$ and (b,d,f) $\Delta =\Omega-J$. Here the point $\omega=\Omega$ corresponds to the flat band in Fig.\,2(a). The parameters are same as that in Fig.\,2 except for $\gamma = 10^{-3}J$.}
	\label{fig3}
\end{figure}

Hybrid interference between the optical and mechanical modes exists in the proposed bipartite optomechanical lattice, which ultimately induces a new flat band together with the photon and phonon localization.

In the case of 1D array, shown in Fig.\,1(a), transforming to the momentum space, the Hamiltonian becomes
\begin{eqnarray}
\label{eq1}
H({\bf{k}}) = \sum_{\bf{k}}(\Delta a_{{\bf{k}}A}^\dagger a_{{\bf{k}}A}+\Delta a_{{\bf{k}}B}^\dagger a_{{\bf{k}}B} + \Omega b_{{\bf{k}}B}^\dagger b_{{\bf{k}},B} \notag\\+ Jfa_{{\bf{k}}A}^\dagger a_{{\bf{k}}B} + ga_{{\bf{k}}A}^\dagger b_{{\bf{k}}B} + H.c. ),
\end{eqnarray}
by a Fourier transformation $ o_{\bf{k}} = \frac{1}{\sqrt{N}}\sum_{n}e^{-i\bf{k}\cdot \bf{R}_n}o_{n}$~( $o_n$ is an arbitrary operator, $N$ is the number of unit cells of the lattice).
Here $f = 1+e^{ik}$, and we have assumed the lattice constant is identical. Under the condition of $\Delta = \Omega$, the band structure is obtained by diagonalizing the Hamiltonian (\ref{eq1}) and it is given by
\begin{eqnarray}
E_0({\bf k}) &=& \Omega, \nonumber
\\
E_\pm({\bf k}) &=& \Omega \pm\sqrt{J^2f^2({\bf k})+g^2}.
\end{eqnarray}
The eigenvalue $E_0 ({\bf{k}}) = 0$ corresponds to the appearance of flat band, which is clearly exhibited in Fig.\,2(a).
Normally, the lattice geometry in a pure photon or phonon 1D lattice
will not induce the destructive interference, and hence no flat band appears in the normal 1D lattice.
Moreover, Fig.\,2 also shows a gap between the middle band and the up (or down) band, which is holden even when the middle flat band disappears under the condition $\Delta\neq\Omega$.
This gap is induced by the optomechanical interaction and its width is decided by the interaction strength.

It should be noticed that, formally, one may naively think this model is a stub lattice~\cite{Hyrkas2013,Baboux2016,Leykam2017} due to the similar Hamiltonian. However, the hybrid 1D optomechanical lattice has three fundamental distinctions comparing with the stub lattice.
First, the flat bands in the two systems have different physical origins. In the stub lattice, the flat band results from the lattice geometry induced destructive interference. But in the 1D bipartite optomechanical lattice discussed above, the flat band is induced by the hybrid interference between two different species of mode, i.e., the optical and the mechanical modes. Second, in other artificial quantum lattice systems, e.g., the photonic crystal, once the stub lattice is constructed, the energy dispersion is fixed. However, as we mentioned above, the energy dispersion of the  optomechanical lattice can be tuned by adjusting the laser detuning $\Delta$. So, as illustrated in Fig. 2, the flat band here can be changed into a dispersive band, and vice versa. Finally, featuring the optomechanical system, the bipartite optomechanical lattice is a one-dimensional lattice. It is contrary to the stub lattice, which is a quasi-one-dimensional lattice.

Different from the previous flat band in the lattice with special geometry structure (e.g., 2D Lieb lattice), this flat band is induced by the photon-phonon hybrid interference between transitions $a_A\leftrightarrow a_B$ and $b_B\leftrightarrow a_B$, which does not exist in the pure photon (or phonon) 1D lattice. This ultimately leads to the result that it has distinguishable photon-phonon-localization property charactered by the special LDOS pattern (see Figs.\,3 and 4).

\begin{figure}[h]
	\centering
	\includegraphics[width = 8.5cm]{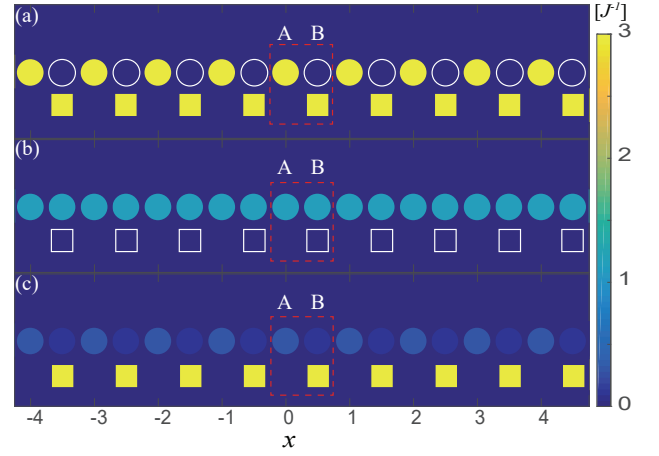}
	\caption{(Color online) (a,b) correspond to the case of $\omega = \Omega$, $\omega = \Omega-J$, respectively, under $\Delta = \Omega$. And (c) describes the LDOS pattern at $\omega = \Omega$, $\Delta = \Omega$. The cycles denote the photonic LDOS pattern and the rectangles denote phononic LDOS pattern. (a) shows the special photon-phonon flat band localization, while the localization does not exist in (b) at $\omega = \Omega -J$ under $\Delta = \Omega$. (c) shows excitations are dispersive, meaning the disappearance of the special localization. Other parameters are same as that in Fig.\,3 except for $\kappa = 0.1J$.}
	\label{fig4}
\end{figure}

\begin{figure}[h!]
	\centering
	\includegraphics[width = 8.5cm]{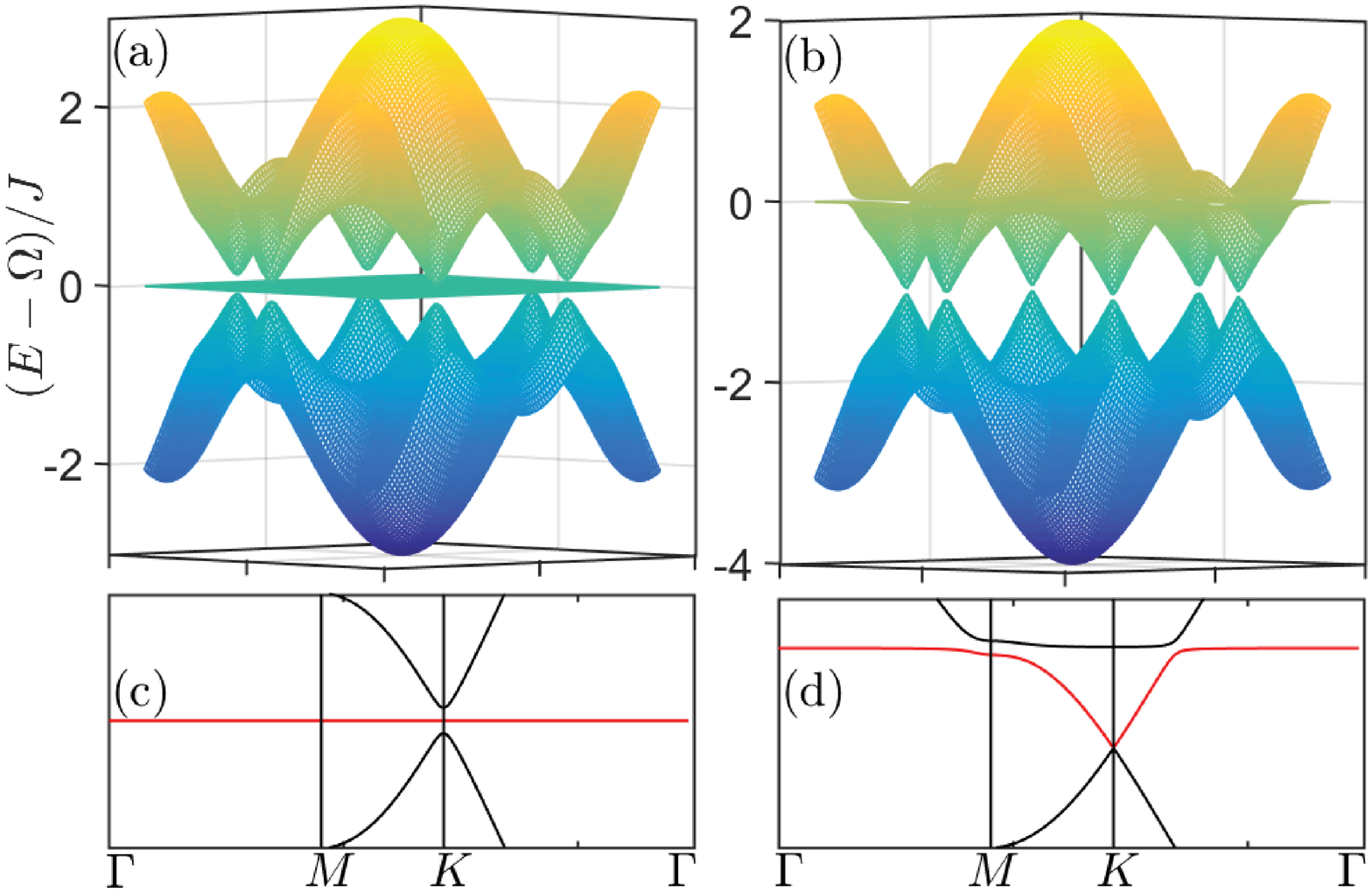}
	\caption{(Color online) (Color online) Energy structure of 2D optomechanical honeycomb lattice under the condition (a,c) $\Delta = \Omega$, (b,d) $\Delta = \Omega-J$. The system parameters are same as that in Fig.\,2.}
	\label{fig5}
\end{figure}

Specifically, the photonic LDOS of sites A, B, and the phononic LDOS of site B are
\begin{eqnarray}
\label{equation9}
\rho_{O}(\omega;j,1) = \frac{-2}{{N}}{\rm{Im}}\sum_{{k}}\frac{(t_1t_2-g^2)}{(t_1t_2-g^2)t_1-J^2|f({{k}})|^2t_2}, \nonumber
\\
\rho_{O}(\omega;j,2) = \frac{-2}{{N}}{\rm{Im}}\sum_{{k}}\frac{t_1t_2}{(t_1t_2-g^2)t_1-J^2|f({{k}})|^2t_2},
\\
\rho_{M}(\omega;j,2) = \frac{-2}{{N}}{\rm{Im}}\sum_{{k}}\frac{t_1^2-J|f({{k}})|^2}{(t_1t_2-g^2)t_1-J^2|f({{k}})|^2t_2}, \nonumber
\end{eqnarray}
where $t_1 = \omega-\Delta+i\frac {\kappa}{2}$, $t_2 = \omega-\Omega+i\frac{\gamma}{2}$, and $\kappa$~($\gamma$) is the dissipation of the optical~(mechanical) mode.
Here the subscripts $O$ and $M$ denote the photon and phonon, respectively. It is shown from Figs.\,3 and 4 that, when the system energy is at the flat band [corresponding to $\omega=\Omega$ in Figs.\,3(a,c,e)] under the condition $\Delta=\Omega$, photons are only localized in the optical sublattice (i.e., A-sites), while phonons are only localized in the optomechanical sublattice (i.e., B-sites). This special LDOS pattern is detectable experimentally by probing the photon and phonon excitations in the lattices, and it offers a simple method to prove the emergence of this new flat band in our model. When the resonant condition $\Delta=\Omega$ is violated, the hybrid photon-phonon-interference is destroyed, leading that the flat band localization disappear [see Figs.\,2(b), 3(b,d,f), and 4(c)]. This demonstrates that the flat band localization in our model is controllable via adjusting the frequency of driving laser. Otherwise, it can be seen that the localization would not emerge when the system energy is not at $\omega = \Omega$, even if the condition $\Delta = \Omega$ is satisfied, as shown in Fig.\,4(b).

\subsection{Flat band localization in two-dimension optomechanical lattice.}

In principle, the presented flat band localization is general and it also could be realized in a two-dimension bipartite optomechanical lattice.
Here we choose the 2D honeycomb and Lieb lattices as the examples. Now the lattice periodicity leads to $f({\bf{k}}) = 1+e^{i{\bf{k}}\cdot{\bf{a}}_1 } + e^{i{\bf{k}}\cdot{\bf{a}}_2}$ [with the basis vector ${\bf a}_1 =(1,0)$, ${\bf a}_2 =(\frac{1}{2}, \frac{\sqrt{3}}{2})$] for the honeycomb lattice, and $f_{j}({\bf{k}})=1+e^{i{\bf{k}}\cdot {\bf{a}}_j}$, $j=1,2$ [with ${\bf{a}}_{1} = (1, 0)$, ${\bf {a}}_2 = (0, 1)$] for the Lieb lattice.

\begin{figure}[h]
	\centering
	\includegraphics[width = 8.5cm]{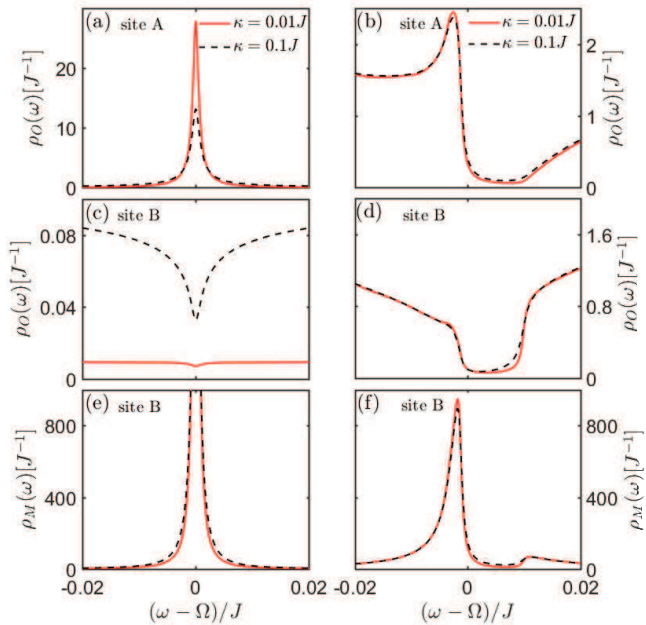}
	\caption{(Color online) The LDOS of 2D optomechanical honeycomb lattice for different decay rate $\kappa$. The local photon or phonon DOS of sites A and B under the condition (a,c,e) $\Delta=\Omega$ and (b,d,f) $\Delta =\Omega-J$. Here the point $\omega=\Omega$ corresponds to the flat band in Figs.\,5(a,c). The parameters are same as that in Fig.\,5 except for $\gamma = 10^{-3}J$.}
	\label{fig6}
\end{figure}

\begin{figure}
	\centering
	\includegraphics[width = 8.5cm]{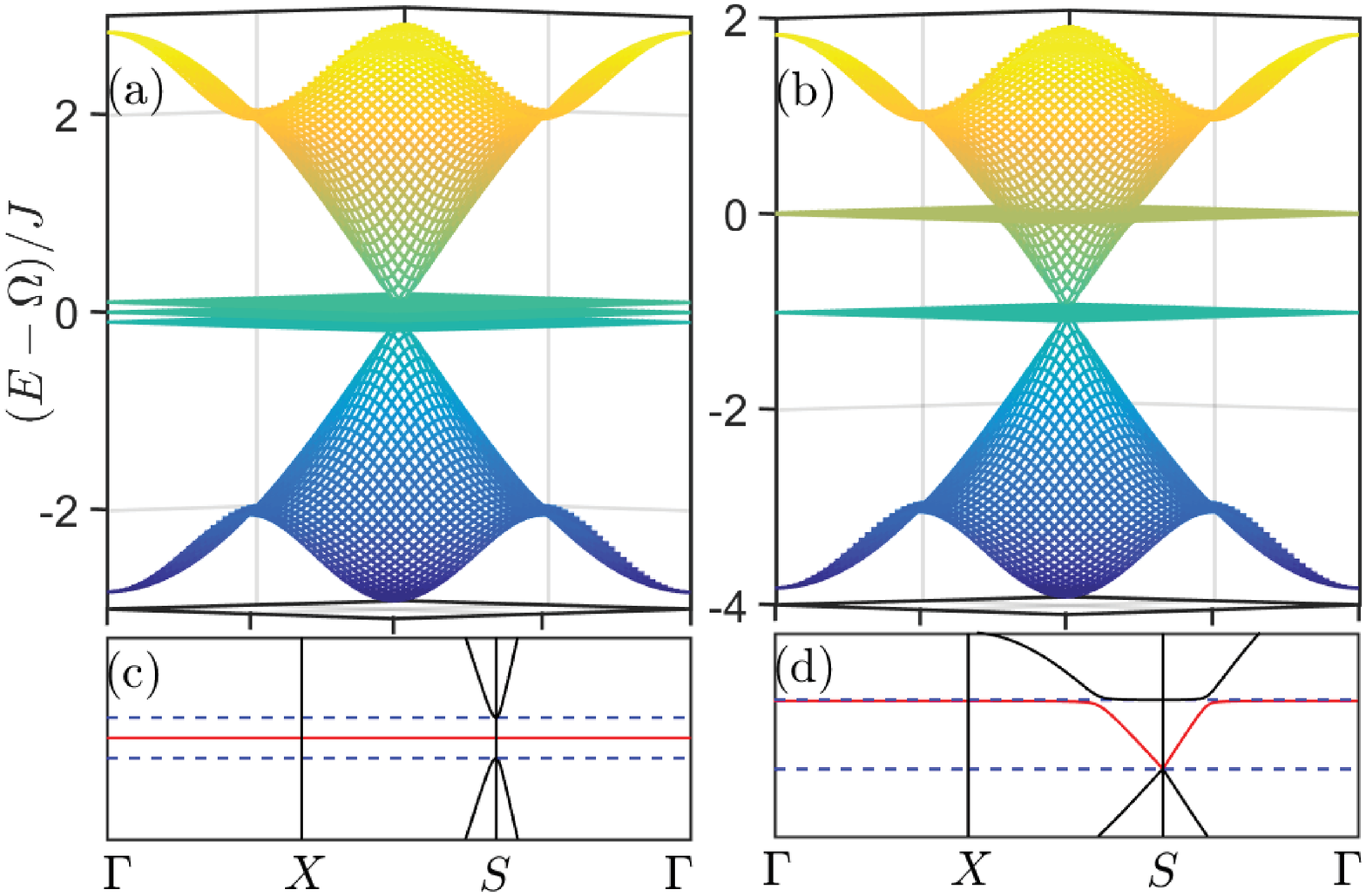}
	\caption{(Color online) Energy structure of 2D optomechanical Lieb lattice under the condition (a,c) $\Delta = \Omega$, (b,d) $\Delta = \Omega-J$. The system parameters are same as that in Fig.\,2.}
	\label{fig7}
\end{figure}

\begin{figure}
	\centering
	\includegraphics[width = 8.5cm]{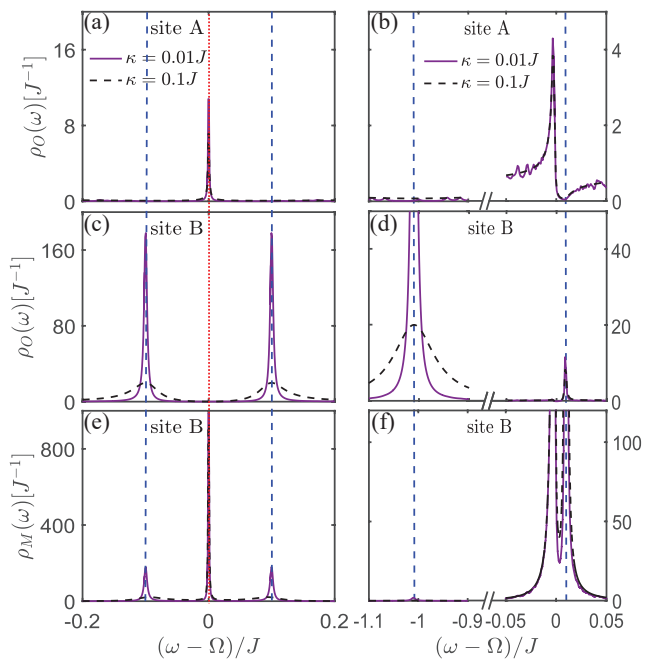}
	\caption{(Color online) The LDOS of 2D optomechanical Lieb lattice for different decay rate $\kappa$. The local photon or phonon DOS of sites A, B and C under the condition (a,c,e) $\Delta=\Omega$ and (b,d,f) $\Delta =\Omega-J$. Here the points $\omega=\Omega$ (indicated by red dashed-dotted line) and $\omega=\Omega\pm g$ (indicated by blue dotted lines) correspond to the middle and the up (or down) flat bands in Figs.\,7(a,c), respectively. Note that, sites B and C have same photon and phonon localization properties. The parameters are same as that in Fig.\,7 except for $\gamma = 10^{-3}J$.}
	\label{fig8}
\end{figure}

In Figs.\,5-8, we plot the energy structure and the LDOS pattern of the hybrid honeycomb and Lieb optomechanical lattices by numerically solving system Hamiltonian.
Firstly, the flat bands are exhibited under the resonant condition $\Delta=\Omega$, as shown in Figs.\,5(a,c) and the middle flat band in Figs.\,7(a,c).
Note that the lattice geometry in a normal honeycomb lattice will not induce the destructive interference, and hence no flat band appears in the normal photon (or phonon) honeycomb lattice.
Even for the normal Lieb lattice, there only is one flat band induced by the destructive interference resulted from its lattice geometry, and it will be holden when the geometry structure is not changed.

Secondly, similar as the case of 1D lattice, the flat band and the middle flat band, respectively, appearing in the optomechanical honeycomb and Lieb lattices are induced by the hybrid photon-phonon-interference.
Because they corresponds to the same photon-phonon-localization pattern shown in the 1D optomechanical lattice, i.e., photons are only localized in the optical sublattice and phonons are localized in the optomechanical sublattice. This is quite different from the flat band localization induced by the destructive interference resulted from Lieb geometry, i.e., the excitations (photon or phonon) are only localized in the sublattice including sites $B$ and $C$. This can be seen more clearly by comparing the points $\omega=\Omega$ (corresponding to the new flat band localization) and $\omega=\Omega\pm g$ (corresponding to the flat band localization in normal Lieb lattice) of Figs.\,8(a,c,e).

\begin{figure}
	\centering
	\includegraphics[width = 8.5cm]{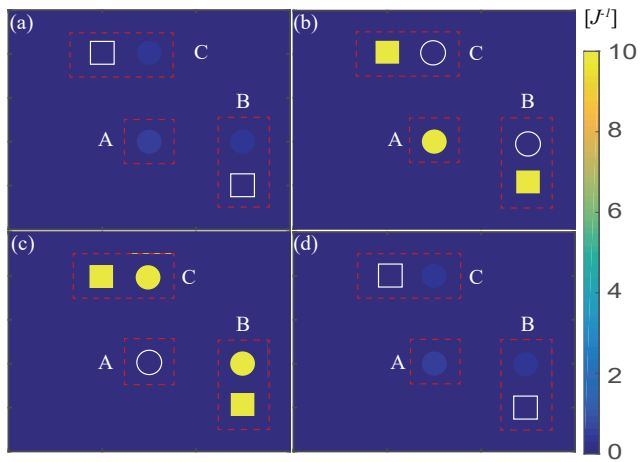}
	\caption{(Color online) The LDOS pattern in real space of the unit cell of Lieb lattice when (a)$\omega = \Omega-J$, (b) $\omega = \Omega$, (c) $\omega = \Omega + g$ and (d) $\omega = \Omega + J$ at $\Delta = \Omega$. The cycles denote the photonic LDOS pattern and the rectangles denote phononic LDOS pattern. Flat band localization does not exist in (a,d). (b) shows the special photon-phonon localization, while (b) shows the localization resulted from its geometry. Other parameters are same as that in Fig.\,3 except for $\kappa = 0.1J$.}
	\label{fig9}
\end{figure}

Lastly, Figs.\,5-8 also show that the hybrid-interference-induced flat band localizations in 2D optomechanical lattices can be controlled by tuning the driving frequency applied in the optomechanical sites (i.e., changing $\Delta$). This also can not be applied into the flat band localizations in the normal Lieb lattice induced by its special geometry structure, as shown in Figs.\,7(b,d) and Figs.\,8(b,d,f).

In addition, Fig.\,9 plots the LDOS pattern in real space. It can been seen that excitations are dispersive, meaning that the localization does not exist for the case of $\omega = \Omega \pm J$, as shown in Fig.\,9(a,d). And Fig.\,9(b,c) show two flat band localizations resulted from distinct origins, which corresponding to $\omega = \Omega$, and $\omega = \Omega + g$. Photons are localized at optical sublattice (i.e., A-sites), and phonons are localized at optomechanical sublattice (i.e., B,C-sites) at $\omega = \Omega$; i.e., the photon-phonon flat band localization in hybrid Lieb lattice. And Fig.\,9(c) shows the intrinsic flat-band localization due to its geometry, in which photons and phonons are localized at optomechanical sublattices (i.e., B, C-sites).

\section{Conclusion}

We have investigated the flat band localization in the bipartite optomechanical lattice both in the cases of 1D and 2D, including a pure optical sublattice and an optomechanical sublattice.
We shown that a new flat band together with a special photon-phonon-localization property is exhibited under the optimal photon-phonon-resonant condition i.e., $\Delta=\Omega$.
This leads to the results that the present flat band localization is detectable experimentally, and can be easily controlled by tuning the frequency of driving laser applied into the optomechanical sublattice.
This study might inspire further explorations regarding the connection bewteen cavity optomechanics and the many-body physics.


\end{document}